\documentclass[12pt]{extarticle}
\usepackage{setspace}

\usepackage{setspace}
\usepackage[T1]{fontenc}
\usepackage{times}
\usepackage{palatino} 
\usepackage{lmodern}
\usepackage[latin1]{inputenc}
\usepackage{epsfig}
\usepackage[english]{babel}
\usepackage{color}
\usepackage{graphicx}
\usepackage{dcolumn}
\usepackage{moreverb}
\usepackage{amsmath,amssymb,amsfonts}
\usepackage[all]{xy}

\begin{document}
\numberwithin{equation}{section}
\newcommand{\boxedeqn}[1]{%
  \[\fbox{%
      \addtolength{\linewidth}{-2\fboxsep}%
      \addtolength{\linewidth}{-2\fboxrule}%
      \begin{minipage}{\linewidth}%
      \begin{equation}#1\end{equation}%
      \end{minipage}%
    }\]%
}


\newsavebox{\fmbox}
\newenvironment{fmpage}[1]
     {\begin{lrbox}{\fmbox}\begin{minipage}{#1}}
     {\end{minipage}\end{lrbox}\fbox{\usebox{\fmbox}}}

\raggedbottom
\onecolumn

\parindent 8pt
\parskip 10pt
\baselineskip 16pt
\noindent\title*{{\LARGE{\textbf{Quadratic algebra structure and spectrum of a new superintegrable system in N-dimension}}}}
\newline
\newline
Md. Fazlul Hoque, Ian Marquette and Yao-Zhong Zhang
\newline
School of Mathematics and Physics, The University of Queensland, Brisbane, QLD 4072, Australia
\newline
E-mail: m.hoque@uq.edu.au; i.marquette@uq.edu.au; yzz@maths.uq.edu.au
\newline
\newline

\begin{abstract}
We introduce a new superintegrable Kepler-Coulomb system with non-central terms in $N$-dimensional Euclidean space. We show this system is multiseparable and allows separation of variables in hyperspherical and hyperparabolic coordinates. We present the wave function in terms of special functions. We give a algebraic derivation of spectrum of the superintegrable system. We show how the $so(N+1)$ symmetry algebra of the $N$-dimensional Kepler-Coulomb system is deformed to a quadratic algebra with only 3 generators and structure constants involving a Casimir operator of $so(N-1)$ Lie algebra. We construct the quadratic algebra and the Casimir operator. We show this algebra can be realized in terms of deformed oscillator and obtain the structure function which yields the energy spectrum.
\end{abstract}

\section{Introduction}
Superintegrable systems form a fundamental part of mathematical theories and modern physics such as quantum chemistry and nuclear physics. They possess many properties in particular analytic and algebraic solvability. Moreover, they have connections to special functions, (exceptional) orthogonal polynomials and Painleve transcendents. Though it has much deeper historical roots, the modern theory of superintegrability was only started 45 years ago \cite{Win1}. A systematic classification of maximally superintegrable systems is now complete for 2 and 3 dimensional Hamiltonians on conformally flat spaces. The classification in higher dimensions and with higher order integrals of motion is much more complicated. In lower dimensions much work has been done for systems involving spins, magnetic fields and monopoles \cite{Win2}. We refer the reader to this review paper for an extended list of references, description of the properties, definitions of superintegrabilty and symmetry algebra in classical and quantum mechanics. One important property of such systems is that they possess non-abelian symmetry algebras generated by integrals of motion. They can be embedded in non-invariance algebras involving non-commuting operators. These symmetry algebras are in general finitely generated polynomial algebras and only exceptionally finite dimensional Lie algebras. The most known examples whose symmetry algebras are Lie algebras generated by integrals of motion are N-dimensional hydrogen atom and harmonic oscillator. See \cite{Foc1,Bar1,Sud1,Ban1,Lou1,Ras1} for systems with $so(N+1)$ symmetry and \cite{Jau1,Bak1,Lou2,Bar1,Hwa1} for those with $su(N)$ symmetry.

Quadratic algebras have been used to provide algebraic derivation of the energy spectrum of superintegrable systems such as the Hartmann system that models the Benzene molecule \cite{Gra1}. A systematic approach for 2D superintegrable systems with quadratic algebra involving three generators was proposed in \cite{Das1}. This method is based on the construction of the Casimir operators and the realization of the quadratic algebra as deformed oscillator. It has recently been generalized to 2D superintegrable systems with cubic, quartic and more generally polynomial algebras \cite{Isa1}. In some cases degeneracy patterns for the energy level are non-trivial and one needs to consider a union of finite dimensional unitary representations to obtain the correct total degeneracies. It has been pointed out how the method can be adapted to study 3D,4D,5D and 8D superintegrable systems \cite{Mar1,Mar2,Das2}. However, the generalization of this approach to N-dimensional superintegrable systems is an unexplored subject. Higher-dimensional superintegrable systems often lead to higher rank polynomial algebras and the structure of these algebras is unknown. The purpose of this paper is to show how we can provide an algebraic derivation of the complete energy spectrum and the total number degeneracies of the $N$-dimensional superintegrable Kepler-Coulomb system with non-central terms. It is based on the quadratic algebra symmetry of the system with structure constant involving Casimir operator of $so(N-1)$ Lie algebra. This is a first step in the study of the polynomial algebra approach to general $N$-dimensional systems. 

The structure of the paper is as follows: In Section 2, we present a new superintegrable Hamiltonian system in $N$-dimensional Euclidean space and show that its Schrodinger wave function is multi-separable in hyperspherical and hyperparabolic coordinates. We present the wave function in terms of special functions and obtain its energy spectrum. In Section 3, we give an algebraic derivation of the energy spectrum of the system. We construct the quadratic symmetry algebra and its Casimir operators. We investigate the realization of the quadratic algebra in terms of deformed oscillator algebra of Daskaloyannis \cite{Das1} and obtain the structure functions which yield the energy spectrum. Finally, in Section 4, we present some discussions with a few remarks on the physical and mathematical relevance of these algebras.

\section{New quantum superintegrable system and separation of variables}
Let us consider the following $N$-dimensional superintegrable Kepler-Coulomb system with non central terms 
\begin{equation}
H=\frac{1}{2}p^{2}-\frac{c_{0}}{r}+\frac{c_{1}}{r(r+x_{N})}+\frac{c_{2}}{r(r-x_{N})}. \label{hamil}
\end{equation}
This system is a generalization of the 3D system that appears in the classification of quadratically superintegrable systems on three dimensional Euclidean space \cite{Eva1,Kib1}. The 3D system has been considered using the method of separation of variables and various results obtained. In this section we apply separation of variables to (~\ref{hamil})
.

\subsection{Hyperspherical Coordinates}
The $N$-dimensional hyperspherical coordinates are given by 
\begin{eqnarray}
&x_{1}&=r sin(\phi_{N-1})sin(\Phi_{N-2})\cdots sin(\Phi_{1}),
\nonumber\\& x_{2}&=r sin(\phi_{N-1})sin(\Phi_{N-2})\cdots cos(\Phi_{1}),
\nonumber\\&...&
\nonumber\\&...&
\nonumber\\&  x_{N-1}&=rsin(\phi_{N-1})cos(\Phi_{N-2}),
\nonumber\\&  x_{N}&=rcos(\phi_{N-1}),
\end{eqnarray}
where the $N$ $x_{i}$'s are Cartesian coordinates in the hyperspherical coordinates, $\{\phi_{1},..., \phi_{N-1}\}$ are the hyperspherical angles and $r$ is the hyperradius. The Schrodinger equation $H\psi=E\psi$ in $N$-dimensional hyperspherical coordinates can be expressed as
\begin{eqnarray}
&[&\frac{\partial^2}{\partial r^2}+\frac{N-1}{r}\frac{\partial}{\partial r}-\frac{1}{r^2}\Lambda^2(N)+\frac{2c'_{0}}{r}-\frac{2c'_{1}}{r^2(1+\cos\phi_{N-1})} \nonumber\\&&\qquad-\frac{2c'_{2}}{r^2(1-\cos\phi_{N-1})}+2E']\psi(r,\Omega)=0 \label{schro}
\end{eqnarray}
where $c'_{0}=\frac{c_{0}}{\hbar{^2}}$, $c'_{1}=\frac{c_{1}}{\hbar{^2}}$, $c'_{2}=\frac{c_{2}}{\hbar{^2}}$ and $E'=\frac{E}{\hbar^{2}}$; and $\Lambda^2(N)$ is the grand angular momentum operator which satisfies the recursive formula
\begin{eqnarray}
-\Lambda^2(N)= \frac{\partial^2}{\partial\phi^2_{N-1}}+(N-2)cot(\phi_{N-1})\frac{\partial}{\partial\phi_{N-1}}-\frac{\Lambda^2(N-1)}{sin^2(\phi_{N-1})},
\end{eqnarray}
valid for all $N$ and $\Lambda^2(1)=0.$ The separation of the radial and angular parts of Eq.(~\ref{schro})
\begin{equation}
\psi(r,\Omega)=R(r)y(\Omega_{N-1}).
\end{equation}
 gives rise to
\begin{eqnarray}
&& \frac{d^2R}{dr^2} +\frac{N-1}{r}\frac{dR}{dr}+(\frac{2c'_{0}}{r}+2E'-\frac{A}{r^2})R=0,\label{rad1} 
\\&&[\Lambda^2(N)+\frac{2c'_{1}}{1+\cos(\phi_{N-1})}+\frac{2c'_{2}}{1-\cos(\phi_{N-1})}-A]y(\Omega_{N-1})=0,\label{an1}
\end{eqnarray}
where $A$ is the separation constant  and $N>1$. Again we may separate the variables of Eq.(~\ref{an1}) \cite{Sae1,Ave1}
\begin{eqnarray}
y(\Omega_{N-1})=\Theta(\phi_{N-1})y(\Omega_{N-2}) 
\end{eqnarray}
we obtain
\begin{eqnarray}
&[&\frac{\partial^2}{\partial\phi_{N-1}^2}+(N-2)\cot(\phi_{N-1})\frac{\partial}{\partial\phi_{N-1}}-\frac{2c'_{1}}{1+\cos\phi_{N-1}}-\frac{2c'_{2}}{1-\cos\phi_{N-1}} \nonumber\\&&\qquad+A-\frac{I_{N-2}(I_{N-2}+N-3)}{\sin^2\phi_{N-1}}]\Theta(\phi_{N-1})=0\label{an2}
\end{eqnarray}
and
\begin{eqnarray}
[\Lambda^2(N-1)-I_{N-2}(I_{N-2}+N-3)]y(\Omega_{N-2})=0, \qquad(N>2)\label{an3}
\end{eqnarray}
where $I_{N-2}(I_{N-2}+N-3)$ is the separation constant and $I_{N-2}\in \mathbb{Z}$.  
The solution of the Eq.(~\ref{an3}) is obtained recursively in $N$.
 
We now turn to Eq.(~\ref{an2}), which can be converted, by setting $z=cos(\phi_{N-1})$ and then $\Theta(z)=(1+z)^{a}(1-z)^{b} f(z)$, to
\begin{eqnarray}
&&(1-z^2)f''(z)+\{2a-2b-(2a+2b+N-1)z\}f'(z)\nonumber \\&&+\{A-(a+b)(a+b+N-2)\}f(z)=0\label{an4}
\end{eqnarray}
where, $2a=\delta_{1}+I_{N-2}$, \quad  $2b=\delta_{2}+I_{N-2}$ and
\begin{eqnarray}
\delta_{i}=\{\sqrt{(I_{N-2}+\frac{N-3}{2})^2+4c_{i}}-\frac{N-3}{2}\}-I_{N-2},\quad i=1, 2. \label{prova2}
 \end{eqnarray}
Comparing Eq.(~\ref{an4}) with the Jacobi differential equation  
\begin{equation}
(1-x^{2})y''+\{\beta-\alpha-(\alpha+\beta+2)x\}y'+\lambda(\lambda+\alpha+\beta+1)y=0,
\end{equation}
we obtain the  separation constant 
\begin{equation}
A=(l+\frac{\delta_{1}+\delta_{2}}{2})(l+\frac{\delta_{1}+\delta_{2}}{2}+N-2)\label{con1} 
\end{equation}
with $l=\lambda+I_{N-2}$. 
Hence the solutions of Eq.(~\ref{an4}) are given in terms of the Jacobi polynomials as
\begin{eqnarray}
&\Theta(\phi_{N-1})&=\Theta_{l I_{N-2}}(\phi_{N-1; \delta_{1}, \delta_{2}})
\nonumber\\&&=F_{l I_{N-2}}(\delta_{1}, \delta_{2}).(1+cos(\phi_{N-1}))^{\frac{(\delta_{1}+I_{N-2})}{2}}.(1-cos(\phi_{N-1}))^{\frac{(\delta_{2}+I_{N-2})}{2}}\nonumber\\&&\qquad\times P^{(\delta_{2}+I_{N-2}, \delta_{1}+I_{N-2})}_{l-I_{N-2}}(cos(\phi_{N-1}))\label{jp1}
\end{eqnarray}
where $P^{(\alpha, \beta)}_{\lambda}$ denotes a Jacobi polynomial and $l\in \mathbb{N}$. The normalization constant $F_{l I_{N-2}}(\delta_{1},\delta_{2})$ in Eq.(~\ref{jp1}) is given by 
\begin{eqnarray}  
&F_{l I_{N-2}}&(\delta_{1},\delta_{2})=\frac{(-1)^{(I_{N-2}-|I_{N-2}|)/2}}{2^{|I_{N-2}|}}\nonumber\\&&\times\sqrt{\frac{(2l+\delta_{1}+\delta_{2}+N-2)(l-|I_{N-2}|)!\Gamma(l+I_{N-2}+\delta_{1}+\delta_{2}+N-2)}{2^{\delta_{1}+\delta_{2}+N-1}\pi\Gamma(l+\delta_{1}+N-2)\Gamma(l+\delta_{2}+N-2)}}.\nonumber\\&&
\end{eqnarray}
\\Let us now turn to the radial equation. Using Eq.(~\ref{con1}) we have
\begin{equation}
\frac{d^2R}{dr^2}+\frac{N-1}{r}\frac{dR}{dr}+[\frac{2c'_{0}}{r}+2E'-\frac{1}{r^2}(l+\frac{\delta_{1}+\delta_{2}}{2})(l+\frac{\delta_{1}+\delta_{2}}{2}+N-2)]R=0.\label{an5}
\end{equation}
Eq.(~\ref{an5}) can be converted, by setting  
 $z=\varepsilon r$, $R(z)=z^{l+\frac{\delta_{1}+\delta_{2}}{2}} e^{-\frac{z}{2}}f(z)$ and $E'=\frac{-\varepsilon^2}{8}$, to
\begin{equation}
z\frac{d^2f(z)}{dz^2}+\{(2l+\delta_{1}+\delta_{2}+N-1)-z\}\frac{df(z)}{dz}-(l+\frac{\delta_{1}+\delta_{2}}{2}+\frac{N-1}{2}-\frac{2c'_{0}}{\varepsilon})f(z)=0\label{an6}
\end{equation}
Set  
\begin{eqnarray}
n=\frac{2c'_{0}}{\varepsilon}-\frac{\delta_{1}+\delta_{2}}{2}-\frac{N-3}{2}.\label{an7}
\end{eqnarray}
Then Eq.(~\ref{an6}) can be written as
\begin{equation}
z\frac{d^2f(z)}{dz^2}+\{(2l+\delta_{1}+\delta_{2}+N-1)-z\}\frac{df(z)}{dz}-(-n+l+1)f(z)=0.\label{an8}
\end{equation}
This is the confluent hypergeometric equation. Hence we can write the solution of Eq.(~\ref{an5}) in terms of the confluent hypergeometric function as  
\begin{eqnarray}
&R(r)&\equiv R_{nl}(r;\delta_{1}, \delta_{2})=F_{nl}(\delta_{1},\delta_{2})(\varepsilon r)^{l+\frac{\delta_{1}+\delta_{2}}{2}}.e^{\frac{-\varepsilon r}{2}}\nonumber\\&&
\quad \times {}_1 F_1(-n+l+1, 2l+\delta_{1}+\delta_{2}+N-1; \varepsilon r).\label{an9}
\end{eqnarray}
The normalization constant $F_{nl}(\delta_{1},\delta_{2})$ from the above relation is given by   
\begin{eqnarray}
&F_{nl}(\delta_{1},\delta_{2})&=\frac{2(-c'_{0})^{3/2}}{(n+\frac{\delta_{1}+\delta_{2}}{2})^2}\frac{1}{\Gamma(2l+\delta_{1}+\delta_{2}+N-1)}\nonumber\\&&\qquad\times\sqrt{\frac{\Gamma(n+l+\delta_{1}+\delta_{2}+N-2))}{(n-l-1)!}}.
\end{eqnarray}
In order to have a discrete spectrum the parameter $n$ needs to be positive integer. From Eq.(~\ref{an7}) 
\begin{equation}
\varepsilon=\frac{2c_{0}}{\hbar^{2}(n+\frac{\delta_{1}+\delta_{2}}{2}+\frac{N-3}{2})}\label{en1}
\end{equation} and hence the energy $E=\frac{-\varepsilon^2 \hbar^2}{8}$ is given by
\begin{equation}
E\equiv E_{n}=-\frac{c^{2}_{0}}{2\hbar^{2}(n+\frac{\delta_{1}+\delta_{2}}{2}+\frac{N-3}{2})^{2}},\qquad n=1,2,3...\label{en1}
\end{equation}
Here $n$ is the principal quantum number.  
\subsection{Hyperparabolic Coordinates}
The $N$-dimensional hyperparabolic coordinates are given by
\begin{eqnarray}
&x_{1}&=\sqrt{\xi \eta} sin(\Phi_{N-2})\cdots sin( \Phi_{1}),
\nonumber\\& x_{2}&=\sqrt{\xi \eta} sin(\Phi_{N-2})\cdots cos( \Phi_{1}),
\nonumber\\&...&
\nonumber\\&...&
\nonumber\\&x_{N-1}&=\sqrt{\xi \eta} cos(\Phi_{N-2}), 
\nonumber\\& x_{N}&=\frac{1}{2}(\xi -\eta), 
\nonumber\\& r&= \frac{\xi+\eta}{2}. 
\end{eqnarray}
where the $N$ $x_{i}$'s are Cartesian coordinates in the hyperparabolic coordinates, $\{\phi_{1},..., \phi_{N-1}\}$ are the hyperparabolic angles and the parabolic coordinates $\xi$, $\eta$ range from $0$ to $\infty$.  The Schrodinger equation $H\psi=E\psi$ in the hyperparabolic coordinates can be written as 
\begin{eqnarray}
&\biggl[-\frac{2}{\xi+\eta}&[\Delta(\xi)+\Delta(\eta)-\frac{\xi+\eta}{4\xi\eta}\Lambda^2(\Omega_{N-1})]-\frac{2c'_{0}}{\xi+\eta}+\frac{2c'_{1}}{\xi(\xi+\eta)} \nonumber\qquad\\&&  +\frac{2c'_{2}}{\eta(\xi+\eta)}\biggr]U(\xi, \eta, \Omega_{N-1})=E'U(\xi, \eta, \Omega_{N-1})\label{fk1}
\end{eqnarray} 
where, $\Lambda^2(N)$ is the grand angular momentum operator defined in the previous subsection and 
\begin{eqnarray*}
&\Delta(\xi)& = \xi^{\frac{-N-3}{2}}\frac{\partial}{\partial \xi}  \xi^{\frac{N-1}{2}}\frac{\partial}{\partial \xi},  \\&\Delta(\eta)& = \eta^{\frac{-N-3}{2}}\frac{\partial}{\partial \eta}  \eta^{\frac{N-1}{2}}\frac{\partial}{\partial \eta},\\& c'_{0}&=\frac{c_{0}}{\hbar{^2}}, \quad c'_{1}=\frac{c_{1}}{\hbar{^2}},\quad c'_{2}=\frac{c_{2}}{\hbar{^2}}\quad and\quad E'=\frac{E}{\hbar^{2}}.
\end{eqnarray*}

The equation can be separated in radial and angular parts by setting 
\begin{eqnarray}
U(\xi, \eta, \Omega_{N-1})=U_{1}(\xi, \eta)y(\Omega_{N-1})
\end{eqnarray}
We obtain two equations 
\begin{eqnarray}
&\biggl[\Delta(\xi)&+\Delta(\eta)-\frac{c'_{1}}{\xi}-\frac{c'_{2}}{\eta}+\frac{E'}{2}\xi+\frac{E'}{2}\eta+c'_{0}-\frac{1}{4\xi}I_{N-2}(I_{N-2}+N-3)\nonumber\\&&-\frac{1}{4\eta}I_{N-2}(I_{N-2}+N-3)\biggr]U_{1}(\xi,\eta)=0\label{fk2}
\end{eqnarray}
and
\begin{equation}
\Lambda^2(\Omega_{N-1})y(\Omega_{N-1})=I_{N-2}(I_{N-2}+N-3)y(\Omega_{N-1})\label{fk3}
\end{equation}
with $I_{N-2}(I_{N-2}+N-3)$ being the general form of the separation constant. The solution of Eq.(~\ref{fk3}) is well-known. By looking for solution of Eq.(~\ref{fk2}) of the form 
\begin{eqnarray}
U_{1}(\xi,\eta)=f_{1}(\xi)f_{2}(\eta),
\end{eqnarray}
we get two coupled equations
\begin{eqnarray}
[\Delta(\xi)-\frac{c'_{1}}{\xi}+\frac{E'}{2}\xi-\frac{1}{4\xi}I_{N-2}(I_{N-2}+N-3)+v_{1}]f_{1}(\xi)=0,\label{fk4}
\end{eqnarray}
\begin{eqnarray}
[\Delta(\eta)-\frac{c'_{2}}{\eta}+\frac{E'}{2}\eta-\frac{1}{4\eta}I_{N-2}(I_{N-2}+N-3)+v_{2}]f_{2}(\eta)=0,\label{fk5}
\end{eqnarray}
where $v_{2}=-v_{1}-c'_{0}$ and $v_{1}$ is the separating constant. Putting $z_{1}=\varepsilon \xi $ in Eq.(~\ref{fk4}), $z_{2}=\varepsilon \eta $ in Eq.(~\ref{fk5}), and $E'=-\varepsilon^2$, these two equations become
\begin{equation}
z_{i}\frac{d^2f_{i}}{dz^2_{i}}+(\delta_{1}+I_{N-2}+\frac{N-1}{2}-z_{i})\frac{df_{i}}{dz_{i}}-(\frac{\delta_{1}+I_{N-2}+\frac{N-1}{2}}{2}-\frac{1}{\varepsilon}v_{i})f_{i}=0\label{fk6}
\end{equation}
with 
\begin{eqnarray*}
 \delta_{i}=\{\sqrt{(I_{N-2}+\frac{N-3}{2})^2+4c'_{i}}-\frac{N-3}{2}\}-I_{N-2},\quad i=1, 2.
\end{eqnarray*}
Let us now denote 
\begin{equation}
n_{i}=-\frac{1}{2}(\delta_{i}+I_{N-2}+\frac{N-1}{2})+\frac{1}{\varepsilon}v_{i}, \quad i=1, 2.\label{fk8}
\end{equation}
Then Eq.(~\ref{fk6}) can be identified with the Laguerre differential equation. Thus we have the normalized wave function
 \begin{eqnarray} 
U(\xi, \eta, \Omega_{N-1})&=&U_{n_{1}n_{2}I_{N-2}}(\xi, \eta, \Omega_{N-1}; \delta_{1},\delta_{2})\nonumber\\&=&\frac{\hbar \varepsilon^2}{\sqrt{-8c_{0}}}f_{n_{1}I_{N-2}}(\xi;\delta_{1})f_{n_{2}I_{N-2}}(\eta;\delta_{2})\frac{e^{iI_{N-2}\Omega_{N-1}}}{\sqrt{2\pi}} 
\end{eqnarray}
where
\begin{eqnarray*}
 &f_{n_{i} I_{N-2}}(t_{i};\delta_{i})& \equiv f_{i}(t_{i})=\frac{1}{\Gamma(I_{N-2}+\delta_{i}+\frac{N-1}{2})}\sqrt{\frac{\Gamma(n_{i}+I_{N-2}+\delta_{i}+\frac{N-1}{2})}{ n_{i}!}} 
 \nonumber\\&&.(\frac{\varepsilon}{2}t_{i})^{(I_{N-2}+\delta_{i})/2}. e^{-\varepsilon t_{i}/4}\times F_{1}(-n_{i}, I_{N-2}+\delta_{i}+\frac{N-1}{2}; \frac{\varepsilon}{2}t_{i})\label{fk7}   
\end{eqnarray*}
with $i=1, 2$ and $ t_{1}\equiv\xi , t_{2}\equiv\eta $. We look for the discrete spectrum and thus $n_{1}$ and $n_{2}$ are both positive integers. 

An expression for the energy of the system in terms of $n_{1}$ and $n_{2}$  can be found by using $E=-\hbar^2 \varepsilon^2$ in Eq.(~\ref{fk8}) to be 
\begin{equation}
E\equiv E_{(n_{1},n_{2})}=\frac{-c^{2}_{0}}{\hbar^{2}\{n_{1}+n_{2}+\frac{1}{2}(\delta_{1}+\delta_{2}+2I_{N-2}+N-1)\}^{2}}.\label{en2}
\end{equation}
We can relate the quantum numbers in (~\ref{en1}) and (~\ref{en2}) by the following relation
\begin{eqnarray}
n_{1}+n_{2}+I_{N-2}=n-1.
\end{eqnarray}
$n_{1}, n_{2}=0, 1, 2,....$. When $I_{N-2}$ is fixed, $p=n_{1}+n_{2}$ provide a degeneracy of $p+1$ for the energy levels. 
\section{Algebraic derivation of the energy spectrum}
In this section we present a algebraic derivation of the energy spectrum for the non-central Kepler-Coulomb system in $N$-dimension. For this purpose we recall some facts about the central Kepler-Coulomb system in the next subsection.

\subsection{Kepler-Coulomb System}
The Hamiltonian of the (central) Kepler-Coulomb system in N-dimensional Euclidean space is given by
\begin{equation}
H=\frac{1}{2}p^{2}-\frac{c_{0}}{r}\label{ham1}
\end{equation}
where $ \vec{r}=(x_{1},x_{2},...,x_{n})$, $\vec{p}=(p_{1},p_{2},...,p_{N})$, $r^{2}=\sum_{i=1}^{N}x_{i}^{2}$ and $p_{i}=-i \hbar \partial_{i}.$
This system has integrals of motion given by the Runge-Lenz vector
\begin{eqnarray}
&M_{j}&=\frac{1}{2} \sum_{i=1}^{N} ( L_{ji}p_{i}-p_{i}L_{ij} )- \frac{c_{0}x_{j}}{r}\\&&=-x_{j}(\frac{1}{2}p^{2}+H)+\sum_{i=1}^{N} x_{i}p_{i}p_{j}-\frac{N-1}{2}i\hbar p_{j}- \frac{c_{0}x_{j}}{r}
\end{eqnarray}\label{rg1}
and the angular momentum 
\begin{equation}
L_{ij}=x_{i}p_{j}-x_{j}p_{i}
\end{equation}
for $i,j=1,2,...,N$.
They commute with the Hamiltonian (~\ref{ham1}),  
\begin{eqnarray*}
[ L_{ij},H]=[ M_{i},H]=0.
\end{eqnarray*}  
The Runge-Lenz vector and angular momentum components generate a Lie algebra isomorphic to $so(N+1)$ for bound states and $so(N,1)$ for scattering state,
\begin{eqnarray*}
 &[L_{ij},L_{kl}]&=i ( \delta_{ik}L_{jl}+ \delta_{jl}L_{ik}-\delta_{il}L_{jk}-\delta_{jk}L_{il})\hbar  
 \\&[M_{i},M_{j}]&=-2i\hbar H L_{ij}, \quad  [M_{k},L_{ij}]=i\hbar (\delta_{ik}M_{j}-\delta_{jk}M_{i}).
\end{eqnarray*}  
An algebraic derivation of the energy spectrum was obtained using a chain of second order Casimir operators (i.e. subalgebra chain $so(N+1) \supset so(N) \supset ... \supset so(2)$ ) to define appropriate quantum numbers \cite{Foc1,Bar1,Sud1,Ban1,Lou1,Ras1}. Another derivation consist in using higher order Casimir operators. This has been performed for the five dimensional hydrogen atom for which the $so(6)$ Casimir operators of order two, three and four, and the related eigenvalues were used to calculate the energy spectrum \cite{Tru1}. The calculation was involved and to our knowledge no such calculation for higher dimensions or even arbitrary dimensions have been done. Let us also remark that the symmetry algebra is not the only algebraic structure of interest and one can use various embedding of the symmetry algebra into a larger one called non-invariance algebra to perform algebraic derivation in particular for $so(4,2)$, $so(7, 4)$ and $sp(8, R)$ \cite{Bar2,Kib2,San1}. 

\subsection{Quadratic Poisson algebra in the non-central Kepler-Coulomb system}
We now consider the non-central Kepler-Coulomb system with Hamiltonian given by (~\ref{hamil}).
This system is superintegrable. The system has the following second order integrals of motion 
\begin{eqnarray}
&A&=\sum_{i<j}^{N}L_{ij}^{2}+ \frac{2 r c_{1}}{r+x_{N}}+\frac{2 r c_{2}}{r-x_{N}}.\label{kf5}
\\&B&=-M_{N}+\frac{c_{1}(r-x_{N})}{r(r+x_{N})}-\frac{c_{2}(r+x_{N})}{r(r-x_{N})}.\label{kf6}
\end{eqnarray}
This can be checked by proving  
\begin{eqnarray*}
\{H,A\}_{p}=\{H,B\}_{p}=0,
\end{eqnarray*}
where $\{,\}_{p}$ is the Poisson bracket defined as $\{X,Y\}_{p}=\sum^{n}_{j=1}(\frac{\partial X}{\partial p_{j}}\frac{\partial Y}{\partial q_{j}}-\frac{\partial X}{\partial q_{j}}\frac{\partial Y}{\partial p_{j}})$. We still have first order integrals of motion 
\begin{eqnarray*}
L_{i j}=x_{i}p_{j}-x_{j}p_{i} \quad for \quad i, j=1, ... , N-1.
\end{eqnarray*}
as $\{H, L_{ij}\}_{p}=0.$
After a long computation, we can show that the integrals of motion generate the quadratic Poisson algebra,
\begin{eqnarray}
&\{A,B\}_{p}&=C,\label{kf1}
\\&\{A,C\}_{p}&=-4 A B + 4 (c_{1}-c_{2})c_{0},\label{kf2}
\\&\{B,C\}_{p}&=2B^{2}-8 H A + 4 J^{2}H+8(c_{1}+c_{2})H-2 c_{0}^{2},\label{kf3}
\end{eqnarray}
where
\begin{eqnarray}
&C&=-\sum_{i,j}^{N}2x_{i}x_{j}p_{i}p_{j}p_{N}+\sum_{i}^{N}[2r^2 p^2_{i}p_{N}+\frac{2c_{0}}{r}x_{i}x_{N}p_{i}-\frac{2c_{1}}{r}x_{i}p_{i}\nonumber\\&&\qquad+\frac{2c_{2}}{r}x_{i}p_{i}] -2c_{0}r p_{N}+\frac{4c_{1}r p_{N}}{r+x_{N}}+ \frac{4c_{2}r p_{N}}{r-x_{N}}.\label{kf7}
\end{eqnarray}
The first order integrals of motion generate a $so(N-1)$ Lie algebra
\begin{eqnarray*}
&\{L_{ij},L_{kl}\}_{p}&= \delta_{ik}L_{jl}+ \delta_{jl}L_{ik}-\delta_{il}L_{jk}-\delta_{jk}L_{il} 
\end{eqnarray*}
for $i, j, k, l=1, ..., N-1.$ Moreover, $\{A, L_{ij}\}_{p}=0=\{B, L_{ij}\}_{p}$. So the full symmetry algebra is a direct sum of the quadratic algebra and $so(N-1)$ Lie algebra.

Thus, $so(N+1)$ symmetry algebra in the central Kepler-Coulomb system is deformed to the quadratic algebra with defined by (~\ref{kf1})-(~\ref{kf3}). Its Casimir operator is given by
\begin{eqnarray}
&&K=C^{2}+4 A B^{2}-8 (c_{1}-c_{2})c_{0}B - 8 H A^{2} + [16 (c_{1}+c_{2})H+8 J^{2}-4 c_{0}^{2}]A.\nonumber\\&&\label{kf4}
\end{eqnarray}
Define
\begin{eqnarray}
J^{2}=\sum_{i<j}^{N-1}L_{ij}^{2}
\end{eqnarray}
$J^2$ is the Casimir operator of the $so(N-1)$ Lie algebra and is also a central element of the Poisson algebra. Other Casimir operators are associated to this $so(N-1)$ Lie algebra. Using the realization for $A$, $B$, $C$ (i.e. (~\ref{kf5}), (~\ref{kf6}) and (~\ref{kf7})), we can show that the Casimir operator (~\ref{kf4}) becomes in terms of the central elements $H$ and $J^2$  
\begin{eqnarray}
K=8(c_{1}-c_{2})^2 H-8(c_{1}+c_{2})c^2_{0}-4c^2_{0}J^2.
\end{eqnarray}
The study of the Poisson algebra and its Casimir operator is important as they will correspond to the lowest order terms in $\hbar$ the quadratic algebra and Casimir operator of the corresponding quantum system.

\subsection{Quadratic algebra in the quantum non-central Kepler-Coulomb system}
We now consider the Hamiltonian of the quantum non-central Kepler-Coulomb system 
\begin{equation}
H=\frac{1}{2}p^{2}-\frac{c_{0}}{r}+\frac{c_{1}}{r(r+x_{N})}+\frac{c_{2}}{r(r-x_{N})}
\end{equation}
Similar to the classical case, the integrals of motion are 
\begin{eqnarray}
&A&=\sum_{i<j}^{N}L_{ij}^{2}+ \frac{2 r c_{1}}{r+x_{N}}+\frac{2 r c_{2}}{r-x_{N}}\label{pr1}
\\&B&=-M_{N}+\frac{c_{1}(r-x_{N})}{r(r+x_{N})}-\frac{c_{2}(r+x_{N})}{r(r-x_{N})}\label{pr2}
\\&J^{2}&=\sum_{i<j}^{N-1}L_{ij}^{2}.
\end{eqnarray}
We still have a set of first order integrals of motion
\begin{eqnarray*}
L_{i j}=x_{i}p_{j}-x_{j}p_{i} \quad for \quad i, j=1, ... , N-1.
\end{eqnarray*}
We can easily verify the commutation relations
\begin{eqnarray*}
[H,A]=[H,B]=[H,J^{2}]=[A,J^{2}]=[B,J^{2}]=[H,L_{ij}]=[L_{ij},J^{2}]=0. 
\end{eqnarray*}
For later convenience we present a diagram representation of the above commutation relations 
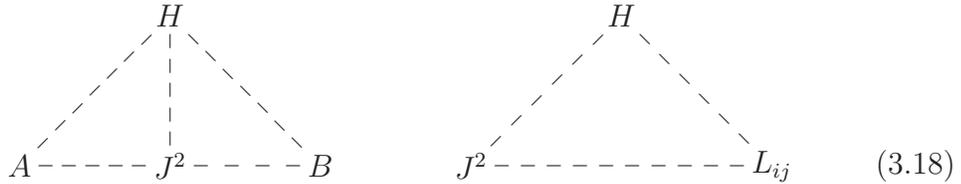
\begin{equation}
\begin{xy}
(0,0)*+{A}="a"; (20,0)*+{J^{2}}="f"; (40,0)*+{B}="b"; (20,20)*+{H}="h";  (60,0)*+{J^{2}}="j"; (100,0)*+{L_{ij}}="l"; (80,20)*+{H}="g"; 
"a";"h"**\dir{--}; 
"h";"b"**\dir{--};
"h";"f"**\dir{--};
"f";"b"**\dir{--};
"f";"a"**\dir{--};
"j";"l"**\dir{--}; 
"l";"g"**\dir{--};
"g";"j"**\dir{--};
\end{xy}
\end{equation}
The left figure shows that $J^2$ is a central element. The right figure illustrates $J^2$ is the Casimir operator of $so(N-1)$ Lie algebra realized by angular momentum $L_{ij}$, $i, j=1, 2,..., N-1$. 
\\After a very tremendous computation, we can show that the integrals of motion close to the quadratic algebra $Q(3)$,
\begin{eqnarray}
&[A,B]&=C,\label{prova5}
\\&[A,C]&=2 \hbar^{2} \{A,B\}+(N-1)(N-3) \hbar^{4} B -4 (c_{1}-c_{2}) \hbar^{2} c_{0},\label{prova6}
\\&[B,C]&=-2 \hbar^{2} B^{2} +8 \hbar^{2} H A -4 \hbar^{2} J^{2} H + (N-1)^{2} \hbar^{4} H \nonumber\\&&\quad-8 \hbar^{2}(c_{1}+c_{2})H + 2 \hbar^{2} c_{0}^{2},\label{prova1}
\end{eqnarray}
where \begin{eqnarray}
&C&=-\sum_{i,j}^{N}2i\hbar x_{i}x_{N}p_{i}p_{j}p_{N}+\sum_{i}^{N}[2 i \hbar r^2 p^2_{i}p_{N}+2\hbar^2 x_{N}p^2_{i}-2N\hbar^2 x_{i} p_{i}p_{N}\nonumber\\&&\quad+ \frac{2i\hbar c_{0}}{r} x_{i}x_{N}p_{i}-\frac{2i\hbar c_{1}}{r}x_{i}p_{i}+\frac{2i\hbar c_{2}}{r}x_{i}p_{i}]+\frac{i\hbar^3}{2}(N-1)^2 p_{N}\nonumber\\&&\quad-2i\hbar c_{0}r p_{N} +\frac{4r c_{1}}{r+x_{N}}i\hbar p_{N}+\frac{4r c_{2}}{r-x_{N}}i\hbar p_{N}+\frac{(N-1)c_{0}}{r}\hbar^2 x_{N}\nonumber\\&&\quad -\frac{(N+1)r+(N-3)x_{N}}{r(r+x_{N})}c_{1}\hbar^2+\frac{(N+1)r-(N-3)x_{N}}{r(r-x_{N})}c_{2}\hbar^2.\label{pr3}
\end{eqnarray}
This quadratic algebra is the quantization of the Poisson algebra in the previous subsection. It can be shown that the Casimir operator is 
\begin{eqnarray}
&K&=C^{2}-2 \hbar^{2} \{A,B^{2}\}+ [ 4 \hbar^{4}-(N-1)(N-3) \hbar^{4}] B^{2} + 8 (c_{1}-c{2})\hbar^{2}c_{0} B\nonumber\\&&+ 8 \hbar^{2} H A^{2} +2[ -4 \hbar^{2} J^{2} H + (N-1)^{2} \hbar^{4} H -8 \hbar^{2} (c_{1}+c_{2})H^{2} +2 \hbar^{2} c_{0}^{2}]A.\nonumber\\&&\label{prova4} 
\end{eqnarray}
By means of the explicit expressions of $A$, $B$, $C$ (i.e. (~\ref{pr1}), (~\ref{pr2}) and (~\ref{pr3})), we can show that the Casimir operator (~\ref{prova4}) becomes in terms of the central elements $H$ and $J^2$   
\begin{eqnarray}
&K&=2(N-3)(N-1)\hbar^{4} H J^{2} -8 \hbar^{2}(c_{1}-c_{2})^{2} H + 4  (N-3) (N-1)\nonumber\\&&\quad .(c_{1}+c_{2}) \hbar^{4} H 
 -\hbar^{6}(N-3)(N-1)^{2} H +4 \hbar^{2} c_{0}^{2} J^{2} \nonumber\\&&\quad+8 \hbar^{2} (c_{1}+c_{2})c_{0}^{2} -2 (N-3) \hbar^{4} c_{0}^{2}.\label{prova3} 
\end{eqnarray}
The first order integrals also generate a $so(N-1)$ Lie algebra as in the classical case
\begin{eqnarray}
[L_{ij},L_{kl}]=i ( \delta_{ik}L_{jl}+ \delta_{jl}L_{ik}-\delta_{il}L_{jk}-\delta_{jk}L_{il})\hbar
\end{eqnarray}
for $i, j, k, l=1, .., N-1.$ Furthermore, $[A,L_{ij}]=0=[B, L_{ij}]$. So the full symmetry algebra is the direct sum of $Q(3)$ and $so(N-1)$ (i.e. $Q(3)\oplus so(N-1)$). A chain of second and higher order Casimir operators are associated with this $so(N-1)$ component. However for the purpose of the algebraic derivation of the spectrum we rely mainly on the quadratic algebra and its Casimir operator.

The quadratic algebra (~\ref{prova1}) can be realized in terms of the deformed oscillator algebra in \cite{Das1}
\begin{eqnarray}
[\aleph,b^{\dagger}]=b^{\dagger},\quad [\aleph,b]=-b,\quad bb^{\dagger}=\Phi (\aleph+1),\quad b^{\dagger} b=\Phi(\aleph)
\end{eqnarray}
where $\aleph $ number operator. We now construct the structure function $\Phi (x)$ by using the Casimir operator (~\ref{prova4}) and the quadratic algebra ((~\ref{prova5}), (~\ref{prova6}),(~\ref{prova1})) as
\begin{eqnarray}
&\Phi (x, u, H)&=3145728 c^2_{0} (c_{1} - c_{2})^2 h^{12} - 196608 h^{12} [8 c^2_{0} (c_{1} + c_{2}) h^2 \nonumber\\&&\quad- 8 (c_{1} - c_{2})^2 h^2 H + 4 c^2_{0} h^2 J^2 - 2 c^2_{0} h^4 (N-3) + 4 (c_{1} + c_{2}) h^4 H \nonumber\\&&\quad .(N-3) (N-1) + 2 h^4 H J^2 (N-3) (N-1) - h^6 H (N-3)\nonumber\\&&\quad .(N-1)^2] \{-1 + 2 (x+u)\}^2 - 1024 h^4 [-128 h^{10} \{2 c^2_{0} h^2\nonumber\\&&\quad - 8 (c_{1} + c_{2}) h^2 H - 4 h^2 H J^2 + h^4 H (N-1)^2\} + 256 h^{14} H \nonumber\\&&\quad .(N-3) (N-1) + 96 h^{10} \{2 c^2_{0} h^2 - 8 (c_{1} + c_{2}) h^2 H\nonumber\\&&\quad  - 4 h^2 H J^2 + h^4 H (N-1)^2\} (N-3) (N-1) - 96 h^{14} H  \nonumber\\&&\quad .(N-3)^2 (N-1)^2] \{-1 + 2 (x+u)\}^2  + 98304 h^{18} H  \nonumber\\&&\quad\{-3 + 2 (x+u)\} \{-1 + 2 (x+u)\}^4 \{1 + 2 (x+u)\} \nonumber\\&&\quad + 512 h^8 [64 h^6 \{2 c^2_{0} h^2  - 8 (c_{1} + c_{2}) h^2 H - 4 h^2 H J^2 \nonumber\\&&\quad+ h^4 H (N-1)^2\} - 128 h^{10} H (N-3)(N-1)] \{-1 \nonumber\\&&\quad  + 2 (x+u)\}^2\{-1 - 12 (x+u) + 12 (x+u)^2\}\label{fk9}
\end{eqnarray}
Here we have also used the expression (~\ref{prova3}) for the Casimir.

A set of quantum number can be defined in same way as \cite{Ras1} with a subalgebra chain for $so(N-1)$ and the related Casimir operators. Thus the eigenvalue of $J^2$ is $\hbar^2 I_{N-2}(I_{N-2}+N-3)$. Also as we act with the structure function $\Phi(x)$ on Fock basis $|n, E>$ with $\aleph|n, E>=n|n, E>$, $H$ in $\Phi(x, u, H)$ can be replaced by $E$. 

To obtain unitary representations we should impose the following three constraints on the structure function :
\begin{equation}
\Phi(p+1, u, E)=0,\quad \Phi(0,u,E)=0,\quad \Phi(x)>0,\quad \forall x>0.\label{pro2}
\end{equation}
where $p$ is a positive integer. These constraints ensure the representations are unitary and finite $(p+1)$-dimensional. The solution of these constrains gives us the energy $E$ and the arbitrary constant $u$. 

From (~\ref{fk9}) and eigenvalues of $J^2$ and $H$, the structure function takes the following factorized form that will greatly simplifies the analysis of finite dimensional unitary representations:
\begin{eqnarray}
&\Phi(x)&=6291456 E\hbar^{18}[x+u-\frac{1-m_{1}-m_{2}}{2}][x+u-\frac{1-m_{1}+m_{2}}{2}] \nonumber\\&&\quad[x+u-\frac{1+m_{1}-m_{2}}{2}][x+u-\frac{1+m_{1}+m_{2}}{2}]\nonumber\\&&\quad .[x+u- (\frac{1}{2}-\frac{c_{0}}{\hbar\sqrt{-2E}}  )] [x+u-( \frac{1}{2}+\frac{c_{0}}{\hbar\sqrt{-2E}}   )] 
\end{eqnarray}
with \begin{eqnarray*}
\hbar^{2}m_{1,2}^{2} =16 c_{1,2}+(4(I_{N-2}(I_{N-2}+N-3)) +(N-3)^{2})\hbar^{2}.
\end{eqnarray*}  
From the condition (~\ref{pro2}), we obtain all possible structure functions and energy spectra, for $\epsilon_{1}=\pm 1$,  $\epsilon_{2}=\pm 1$.
\\Set-1:
\begin{eqnarray}
u=\frac{1}{2}+\frac{c_{0}}{\hbar\sqrt{-2E}},  \qquad E=\frac{-2 c^2_{0}}{h^2 (2 + 2 p +\epsilon_{1} m_{1} +\epsilon_{2} m_{2} )^2}
\end{eqnarray}
and
\begin{eqnarray}
&\Phi(x)&=\frac{786432 c^2_{0} h^{16} x[2 + 2 p +x+ \epsilon_{1} m_{1} +\epsilon_{2} m_{2}] }{(2 + 2 p + \epsilon_{1} m_{1} +\epsilon_{2} m_{2} )^2}  \nonumber\\&&\quad .[ 2 + 2p - 2x+(1 + \epsilon_{1})m_{1}+ (1 + \epsilon_{2}) m_{2} ]\nonumber\\&&\quad .[2x- 2 - 2p +(1 -\epsilon_{1}) m_{1} - (1 +\epsilon_{2})m_{2}  ]  \nonumber\\&&\quad .[2x-2p-2+(1-\epsilon_{1})m_{1}+(1-\epsilon_{2})m_{2}]\nonumber\\&&\quad .[2x-2p-2-(1+\epsilon_{1})m_{1}+(1-\epsilon_{2})m_{2}]. 
 \end{eqnarray}
\\Set-2:
\begin{eqnarray}
u=\frac{1}{2}-\frac{c_{0}}{\hbar\sqrt{-2E}},\qquad E=\frac{-2 c^2_{0}}{h^2 (2 + 2 p +\epsilon_{1} m_{1} +\epsilon_{2} m_{2} )^2}
\end{eqnarray}
and
\begin{eqnarray}
&\Phi(x)&=\frac{786432 c^2_{0} h^{16} x [2+2p-x+\epsilon_{1}m_{1}+\epsilon_{2}m_{2}]}{(2 + 2 p + \epsilon_{1} m_{1} +\epsilon_{2} m_{2} )^2}  \nonumber\\&&\quad .[ 2 + 2p+2x+(1 + \epsilon_{1})m_{1}+ (1 + \epsilon_{2}) m_{2}]\nonumber\\&&\quad .[2 + 2 p + 2x-(1-\epsilon_{1}) m_{1} -(1-\epsilon_{2})m_{2}]
\nonumber\\&&\quad .[2+2p+2x+(1+\epsilon_{1})m_{1}-(1-\epsilon_{2})m_{2}]\nonumber\\&&\quad .[2+2p+2x-(1-\epsilon_{1})m_{1}+(1+\epsilon_{2})m_{2}]. 
 \end{eqnarray}
\\Set-3:
\begin{eqnarray}
u=\frac{1}{2}(1+\epsilon_{1}m_{1}+\epsilon_{2}m_{2}), \qquad  E=\frac{-2 c^2_{0}}{h^2 (2 + 2 p +\epsilon_{1} m_{1} +\epsilon_{2} m_{2} )^2}
\end{eqnarray}
and
\begin{eqnarray}
&\Phi(x)&=\frac{786432 c^2_{0} h^{16}[1+p-x]}{(2 + 2 p + \epsilon_{1} m_{1} +\epsilon_{2} m_{2} )^2}   \nonumber\\&&\quad .[ 2 + 2p+(1 +\epsilon_{1})m_{1}- (1 - \epsilon_{2}) m_{2}]\nonumber\\&&\quad .[2 + 2 p+(1+\epsilon_{1}) m_{1} +(1+\epsilon_{2})m_{2}]
 \nonumber\\&&\quad .[2+2p-(1-\epsilon_{1})m_{1}+(1+\epsilon_{2})m_{2}]\nonumber\\&&\quad .[2x-(1-\epsilon_{1})m_{1}-(1-\epsilon_{2})m_{2}]\nonumber\\&&\quad .[1+p+x+\epsilon_{1}m_{1}+\epsilon_{2}m_{2}].
\end{eqnarray}
The structure functions are positive for the constraints $\varepsilon_{1}=1$\quad $\varepsilon_{2}=1$ and $m_{1}, m_{2}>0$.
Using formula (~\ref{prova2}), we can write $m_{1}$ and $m_{2}$ in terms of $\delta_{1}$, $\delta_{2}$ and $I_{N-2}$ as $m_{1}=\frac{1}{2}(3-2I_{N-2}-N-2\delta_{1})$, $ m_{2}=\frac{1}{2}(3-2I_{N-2}-N-2\delta_{2})$. Making the identification $p=n_{1}+n_{2}$, the energy spectrum becomes (~\ref{en2}).

\section{Conclusion}
One of the main results of this paper is the construction of the quadratic algebra for the N-dimensional non-central Kepler-Coulomb system. We obtain the Casimir operators and derive the structure function of the deformed oscillator realization of the quadratic algebra. The finite dimensional unitary representations of the algebra yield the energy spectrum. We compare our results with those obtained from separation of variables. 
 
Algebra structures appearing in $N$-dimensional superintegrable systems are an unexplored area. More complicated polynomial algebra structures are expected in general and it is non-trivial to generalise the present approach to these cases. Let us mention the possible generalizations to monopole interaction and their dual based on \cite{Mar1,Mar2}. Moreover, the classification of certain families of superintegrable systems  with quadratic integrals of motion in N-dimensional curved spaces have been done and their quadratic algebra structures should be studied \cite{Bal1}. In recent a paper \cite{Dan1} a superintegrable system with spin has been obtained. An algebraic derivation of the spectrum would be of interest.

Let us point out that 2D superintegrable systems and their quadratic algebras have been related to the full Askey scheme of orthogonal polynomials via a contraction process. This illustrates a deep conection between superintegrable systems, orthogonal polynomials and quadratic algebras \cite{Wil1}. The relations between the quadratic algebras of superintegrable systems involving Dunkl operators and special functions have been studied in a series of papers \cite{Vin1,Vin2}. It would be interesting to generalize the results to $N$-dimensional superintegrable systems. 

{\bf Acknowledgements:}
The research of F.H. was supported by International Postgraduate Research Scholarship and Australian Postgraduate Award. I.M. was supported by the Australian Research Council through a Discovery Early Career Researcher Award DE 130101067. YZZ was partially supported by the Australian Research Council, Discovery project DP 110103434 and DP 140101492.

\end{document}